PAPER • OPEN ACCESS

# Properties of piezoceramic materials in high electric field actuator applications

View the article online for updates and enhancements.

# Properties of piezoceramic materials in high electric field actuator applications

Binal P Bruno, Ahmed Raouf Fahmy, Moritz Stürmer, Ulrike Wallrabe and Matthias C Wapler

Laboratory for Microactuators, IMTEK—Department of Microsystems Engineering, University of Freiburg, Georges-Köhler-Allee 102, D-79110, Freiburg, Germany

E-mail: wallrabe@imtek.uni-freiburg.de

## Abstract
In this paper, we compare the performance of 8 PZT ceramics and one PMN-PT material for typical bending actuator applications. This includes the measurement of nonlinear transverse charge coefficient at high electric field strength and related quantities such as the Young's modulus, relative permittivity, coercive field and their temperature dependencies, and the Curie temperature. Most materials show much higher strains than what is expected from the datasheet values. We further study the operating region for fields against the polarization direction in different operating cycles and demonstrate a long-term stable quick re-poling method which increases the operating range of negative-only cycles from 50% of $E_c$ to 66% of $E_c$.

Keywords: piezo ceramics, piezo actuators, PMN-PT, nonlinearity

(Some figures may appear in colour only in the online journal)

## 1. Introduction

Today, piezoceramic materials are used in a wide range of MEMS applications, for example in the field of microsensors [1, 2], ultrasonic actuators [3] and motors [4], micropumps [5], adaptive lenses [6–8], tunable mirrors [9] and tunable optical gratings [10]. As the relevance of piezoceramics is increasing in the field of MEMS, new materials are also being developed according to the needs of the industry. The most recent class of PMN-PT materials, when synthesized in single crystalline form, shows remarkably large piezoelectric constants in comparison to the traditional PZT ceramics [11]. Even though there are MEMS methods for depositing piezo materials [12], discrete assembly of piezo sheets is still dominant.

We can obtain PZT ceramics from a variety of manufacturers with different material parameters and with a thickness ranging from 100 μm to few millimeters. By making piezoceramic sheets thinner than 200 μm, high electric fields can be achieved with relatively low voltages which makes them useful for an increasing number of actuator applications. However, due to the nonlinearity of piezoceramics, their behavior at high electric fields may be different from the specifications in the material datasheets that are usually obtained at small fields and assume a linear behavior [13, 14]. In this limit, the dielectric displacement $D$ and strain $S$ can be expressed in the strain-charge form as a function of the mechanical stress $T$ and electric field $E$ by the piezoelectric linear constituent equations

$$D_m = d_{mi}\, T_i + \epsilon^T_{ik}\, E_k, \quad (1)$$

$$S = s^E_{ij}\, T_i + d_{im}\, E_k. \quad (2)$$

The piezoelectric charge coefficient $d$ and the permittivity at constant stress $\epsilon^T$ describe the response of piezoceramics to the applied electric field.

A number of extensive studies have already been focused on the nonlinear behavior of piezoceramic materials at high electric fields and are summarized in [15], whereas nonlinear constitutive relations for piezoceramic materials can be found in [16]. Wang *et al* measured the nonlinear behavior of *PZT 3203 HD* from *Motorola*, classified as soft PZT ceramic, up to an electric field of 0.15 kV mm$^{-1}$ by measuring the tip deflection and blocking force of cantilever beams using an optical fiber displacement sensor and a force load cell, respectively [17]. The results were compared to nonlinear analytical solutions by Chattaraj *et al* [18] and Yao *et al.* [19]. The behavior of 'soft' piezoelectric ceramics (*PZT5H* from *Morgan Matroc*, *PZT 3203 HD* from *Motorola* and *PK1550* from *Piezo Kinetic Inc.*) at a sinusoidal electric field up to 0.14 kV mm$^{-1}$ was studied by Kugel *et al* using strain gauges [20]. The piezoelectric coefficient of PMN-PT was measured by Taylor *et al* by measuring the change in length of the sample as a function of the electric field up to 0.6 kV mm$^{-1}$ [21, 22]. Furthermore the change in permittivity of the PZT-5H material has been already studied by Sirohi *et al* by measuring the impedance of piezoceramic sheet actuators at electric fields ranging from 0.15 to 0.45 kV mm$^{-1}$ [23] and the temperature dependance of the charge coefficient of PZT-5H from −125 °C to +125 °C (under large electric fields (0.5 kV mm$^{-1}$)) was studied by Wang *et al* using strain gauges glued to the ceramic [24]. In all of these studies, however, the maximum applied electric field is at most 0.6 kV mm$^{-1}$, which is well below the maximum field strengths found in typical actuator applications [25–27]. While some compare different types of piezoelectric ceramics, they do not compare a set of specific materials of different manufacturers and the manufacturers, in turn, do not provide the type, composition or doping of their PZT ceramics.

In this study, we investigate the full nonlinear behavior of the essential material parameters at large field strengths up to 1.5 kV mm$^{-1}$ and the maximum strain of different piezoelectric materials from different manufacturers. We use a typical bending actuator configuration with 100–200 $\mu$m thick piezo films which is similar to the method adopted by Wang *et al* [17]. Our method, however, will use the curvature of the beam rather than the deflection at the tip which we believe gives a more accurate result that is more tolerant to fabrication and alignment uncertainties. We also compare our results to the results in the literature mentioned above. Furthermore, we introduce novel methods to increase the operating electric field limits of the PZT ceramics by quick re-poling and symmetric driving methods. Our measurement approach is not designed according to the DIN EN 50324 standard but instead considers the low frequency and high electric field conditions in which piezo actuators typically operate in MEMS applications. We further studied the temperature dependence of the piezoelectric properties and measured the maximum electric field that can be applied against the direction of polarization of the piezoceramics in different operating cycles and long-term actuations.

The materials include PZT-5H and PZT-5A piezoceramics with varying range of coercive field, charge coefficients and Curie temperature. We also acquired samples of PMN-PT to compare its performance with PZT ceramics. Section 2 describes the materials used in the study, the fabrication of the bending actuator and the measurement of the transverse charge coefficient, relative permittivity and Young's modulus of the piezo-ceramics at high electric fields and high mechanical strains. In section 3, we measure the Curie temperature, the temperature dependence of the charge coefficient and Young's modulus and the long-term stability in different electric operating ranges. The results are discussed in section 4, where we compare various performance figures of different materials, followed by the conclusion of the study in section 5.

## 2. Piezoelectric charge coefficient

In the following section, we describe the piezoceramic materials that we use, the measurement method adopted to evaluate the charge coefficient and the measurement of the auxiliary quantities that we need for the evaluation of the charge coefficient.

When an electric field is applied across a piezoceramic in the direction of polarization (out-of-plane), a positive strain is generated in the same direction resulting in elongation in the polarization direction and contraction in the transverse direction (in-plane). Due to the low strains that are achieved, many MEMS applications use piezo actuators in bending configurations, where two active layers (bimorph) or an active and a passive layer (unimorph) are bonded together. Wang *et al* compared the nonlinearity in the tip deflection of unimorph, bimorph and RAINBOW actuators [17]. Hence, to measure the electromechanical behavior, we fabricated bending beams from piezoceramic sheets and measured their bending profile at different quasi-static electric fields, using an optical profilometer. We compare the measured profile with that of a simulated beam having the same material properties, except for the charge coefficient which is kept at the standard value for a PZT-5H (−274 pm V$^{-1}$). The charge coefficient of the measured beam is then obtained by comparing the measured and simulated curvature coefficients. As COMSOL *Multiphysics* is currently unable to simulate nonlinearity in piezoceramic materials, we assume that we are in the linear regime, such that we can use the linear relationship between the curvature of the beam and the strain to determine the $d_{31}$ for each applied electric field strength. The mechanically nonlinear simulation in COMSOL was not implemented as the effect is significant only in situations involving buckling of the actuator. We verified this by comparing the linear and nonlinear simulation in COMSOL and found deviations in the simulated profile to be in the order on several hundred nanometers. Hence, we also simply use the much faster and more reliable linear simulation in our model. To achieve an accurate simulation, we have to evaluate all the material parameters affecting the simulation model. In the strain-charge form of piezoelectric simulation in COMSOL *Multiphysics*, we need the relative permittivity, density and Young's modulus of the material.

### 2.1. Sample preparation

We obtained the piezoceramic samples from five different manufacturers: *American Piezo, Johnson Matthey, Ekulit GmbH, PI Ceramics* and *TRS Technologies*. The materials are listed in table 1 along with the values of the material parameters from the datasheet. *Ekulit-PZT* and *AP-PZT856* have no material datasheet available. The material name as given in table 1 will be used throughout this paper to denote the materials. The piezoceramic from *Ekulit GmbH* comes in the form of a piezo buzzer where we separated the piezo layer from its metal base by dissolving the glue in *Dichloromethane*. All other materials are supplied in monolayer sheets with electrode layers on both sides. We added a PMN-PT material from *TRS Technologies* to the study to compare its performance to PZT ceramics. According to the mechanical quality factor ($Q_m$) which we measured for the materials whose quality factor was not provided by the manufacturer, all materials except JM-M1334 were soft PZT materials since their quality factor was below 150 and JM-M1334 has a quality factor of 220.

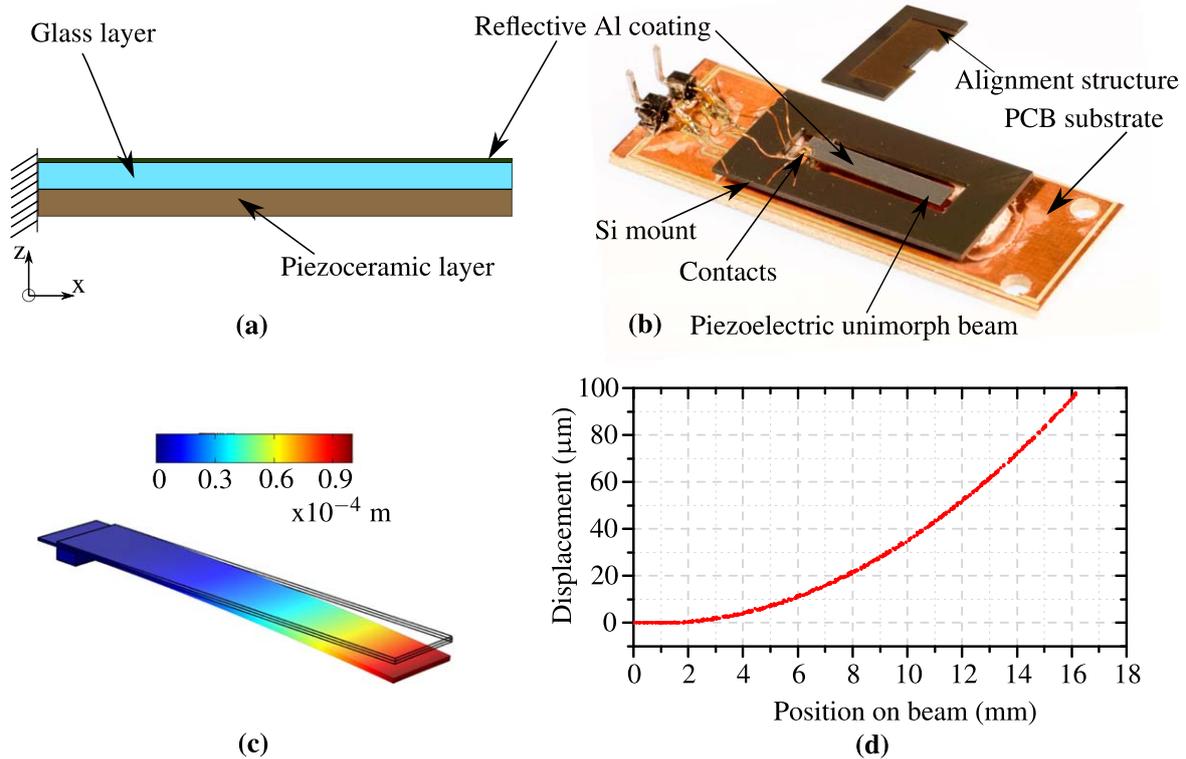

**Figure 1.** (a) Cross-section of the piezoelectric unimorph beam showing the active piezo layer and the passive glass layer coated with aluminum (not to scale). (b) Piezoelectric beam mounted on the silicon structure and contacted to a PCB substrate for easier handling and the alignment structure for gluing the beams. (c) Simulated unimorph beam. (d) An example of bending profile of a simulated unimorph beam.

**Table 1.** The materials used in the study, their manufacturers and the material parameters as given in the datasheet.

| Material | Manufacturer | Density (Kg m$^{-3}$) | $T_c$ (°C) | $\epsilon_r$ | $E_c$ (kV mm$^{-1}$) | $d_{31}$ (pm V$^{-1}$) | Thickness (μm) |
|---|---|---|---|---|---|---|---|
| AP-PZT850 | American piezo | 7600 | 360 | 1900 | — | 175 | 100 |
| AP-PZT856 | American piezo | — | — | — | — | — | 300 |
| EKULIT-PZT | Ekulit GmbH | — | — | — | — | — | 120 |
| JM-M1100 | Johnson Matthey | 8100 | 177 | 4750 | 0.57 | 315 | 120 |
| JM-M1334 | Johnson Matthey | 7900 | 200 | 3500 | 0.615 | 230 | 150 |
| TRS-207HD | TRS technologies | 7950 | 370 | — | 1.6 | 202 | 150 |
| TRS-610HD | TRS technologies | 7950 | 210 | — | 0.78 | 380 | 150 |
| TRS-PMNPT | TRS technologies | — | — | — | 0.45 | 1200 ($d_{33}$) | 205 |
| PI-PIC252 | PI ceramics | 7800 | 350 | 1650 | — | 180 | 110 |

We chose a configuration with a unimorph beam that uses an active layer of piezoelectric material and 100 μm thick borosilicate glass (*D263T-eco Thin Glass* from *Schott AG*) as a passive layer. We evaporated a thin metal layer of 10 nm chromium and 300 nm aluminum on to the glass layer to provide a high-quality optically reflective surface to reliably measure the surface with an optical profilometer (figure 1 (a)) as described in more detail in sections 2.3 and 2.4. The glass and PZT were structured by UV laser and glued using epoxy resin (*HTG-240* from *Resoltech*) . The resin was cured for 8 h at 40 °C and then at 80 °C for 2 h, which resulted in a glass transition temperature of the glue layer ($T_g$) of about

**Table 2.** Measured values of density, relative permittivity, Curie temperature and the coercive field strength for all materials in the study and their datasheet values in parentheses.

| Material | Density | | $\epsilon_r$ | | Curie temperature | | Coercive field @ 25 °C | |
| --- | --- | --- | --- | --- | --- | --- | --- | --- |
| | (kg m$^{-3}$) | | | | $T_c$ (°C) | | $E_c$ (kV mm$^{-1}$) | |
| AP-PZT850 | 7700 ± 200 | (7600) | 1700 ± 100 | (1900) | 280 ± 30 | (360) | 0.71 ± 0.02 | |
| AP-PZT856 | 7800 ± 300 | | 3000 ± 100 | | 210 ± 10 | | 0.85 ± 0.02 | |
| EKULIT-PZT | 7400 ± 500 | | 3300 ± 100 | | 160 ± 2 | | 0.77 ± 0.01 | |
| JM-M1100 | 7900 ± 400 | (8100) | 4700 ± 200 | (4750) | 160 ± 6 | (177) | 0.52 ± 0.01 | (0.57) |
| JM-M1334 | 7900 ± 200 | (7900) | 3400 ± 100 | (3500) | 170 ± 8 | (200) | 0.54 ± 0.01 | (0.615) |
| TRS-207HD | 8000 ± 30 | (7950) | 2300 ± 100 | | 350 ± 10 | (370) | 1.30 ± 0.03 | (1.60) |
| TRS-610HD | 7900 ± 10 | (7950) | 4100 ± 200 | | 250 ± 10 | (210) | 0.59 ± 0.03 | (0.78) |
| TRS-PMNPT | 8200[a] | | 4700 ± 176 | | 100 ± 12 | | 0.33 ± 0.04 | (0.45) |
| PI-PIC252 | 7800 ± 200 | (7800) | 1700 ± 80 | (1650) | 389 ± 19 | (350) | 1.34 ± 0.04 | |

[a] Measured only for one sample because of the limited availability of material. Density of all other materials is averaged over 3 samples.

150 °C. Even though it is possible to achieve higher $T_g$ by increasing the curing temperature of the glue, we chose a low curing temperature as the materials in this study have the Curie temperature in the range of 140 °C–400 °C. We glued the finished unimorph beams to a silicon mount as shown in figure 1 (b) using an alignment structure; both also structured using the UV laser. The beams had a free effective length of 12–15 mm and width of 2–3 mm depending on the size of the sample available.

### 2.2. Relative permittivity

One of the main material parameter affecting our COMSOL simulation is the relative permittivity of the material, which we measured by structuring the piezoceramic samples to a defined electrode surface area using a UV laser and measuring the capacitance across the electrodes. This gives us the $\varepsilon_r$ at constant mechanical stress which is shown in table 2.

To evaluate the value for $\varepsilon_r$ at different electric fields, we applied a sinusoidal signal with different amplitudes in the direction of polarization to a sample single layer of the piezo material of JM-M1334 of known area. The frequency of the signal was kept at 100 mHz in order to measure the voltage and current in a well-defined static regime of the material. We then obtained the high-field relative permittivity at each voltage cycle from the integrated *V–I* relation. The results are discussed in section 2.4.

### 2.3. Young's modulus

The Young's modulus of a material can be measured using static measurement techniques like tensile, bending or torsion tests, but also doing dynamic measurement methods by measuring the resonance frequency of the material. As the forces or strains in static tests may be non-trivial to measure, we adopted the latter by measuring the resonance frequency of the unimorph beams. We excited the beams with low electric fields of about 0.05 kV mm$^{-1}$ over a frequency sweep and recorded the amplitude of the deflection using a high-speed triangulation sensor (*Keyence LK-H022K*) to obtain the resonance frequency. To obtain the Young's modulus from the resonance frequency, we further needed the density of the material. Even though the datasheet of the materials provides us with a value for the density, we also measured it using the traditional water displacement method.

We then modeled the unimorph beams in COMSOL and carried out an eigenfrequency simulation with a parameter sweep over the Young's modulus by scaling the compliance matrix of the material. By taking the inverse of the (1, 1)th value of the compliance matrix corresponding to the measured resonance frequency, we got the Young's modulus of the material, which is given in table 2. The uncertainty in the values of the Young's modulus originates from the variations in the fabrication process of the beams that affect the effective length of the beam, the width, and the thickness and also the uncertainty of the density value. Since PZT is a highly non-linear material, we furthermore measured the Young's modulus for different amplitudes of the displacement by increasing the electric field. To best quantify the result, we use the volume average strain of the PZT at the amplitude of the resonance in the simulations to represent the strain at which the Young's modulus was measured. In figure 2 we see that the measured Young's modulus decreases with increasing strain.

### 2.4. Charge coefficient $d_{31}$

The piezoelectric charge coefficient or piezo modulus ($d_{ij}$) is defined as the ratio of the charge displacement in a certain direction (*i*) to an applied force (*j*), or the strain (*j*) to the applied electric field (*i*). The three independent charge coefficients, $d_{33}$, $d_{31}$, and $d_{15}$ then describe the longitudinal, transverse and shear deformation of polarized ceramics respectively. The measurement of $d_{33}$ and $d_{15}$ is not covered in this study as most bending actuator applications utilize the transverse strain, i.e. the charge coefficient, $d_{31}$.

To evaluate this charge coefficients at different electric fields, we excited the beams with a low-frequency sinusoidal signal of 1 Hz and different amplitudes in the direction of polarization. We then measured the bending profile of the beams dynamically using an optical profilometer equipped with a chromatic confocal distance sensor with a vertical and lateral resolution of 75 nm and 4 μm, respectively. We used a

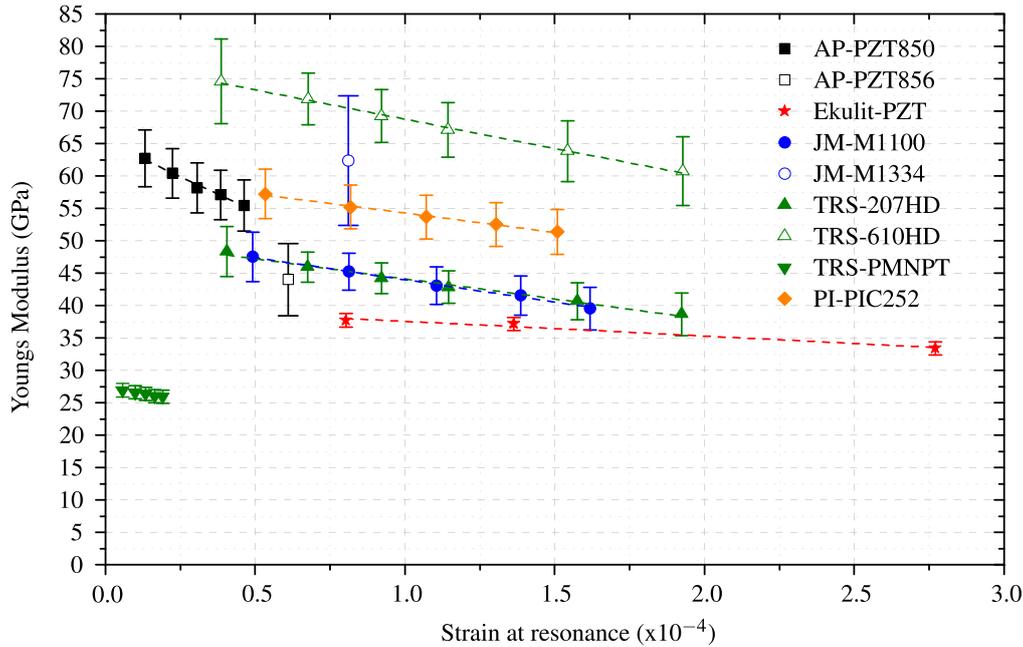

**Figure 2.** The measured Young's modulus of different materials as a function of the strain inside the material caused by the displacement at the resonance frequency.

100 $\mu$m raster and averaged the voltage-dependent displacement over five voltage cycles to minimize the error and extracted the bending profile at the maximum of the corresponding voltage cycle and fitted a second-order polynomial function

$$f(x) = a + bx + c_{meas}x^2 \quad (3)$$

on segments of 2 mm along the length of the beam to obtain the local curvature coefficient. To avoid edge effects, we did not consider the outer 200 $\mu$m of the beam.

To determine the charge coefficient, we modeled the beam in a COMSOL multiphysics FEM simulation. As we could only measure one component of the compliance matrix and the permittivity matrix, we scaled the matrices of the built-in PZT-5H material with our measured values. The coupling matrix was kept at the default value with a charge coefficient of $-274$ pm V$^{-1}$ ($d_{31}{}^{sim}$). We simulated the bending profile (figures 1(c) and (d)) for electric fields as in the physical measurement and fitted the simulated curvature with a second-order polynomial function, extracting the simulated curvature coefficient ($c_{sim}$) in the same method as in the measurement. Since the ratio of measured curvature ($c_{meas}$) to the simulated curvature ($c_{sim}$) at the same electric field $E$, is equal to the ratio of charge coefficient $d_{31}$ to the charge coefficient used in the simulation ($d_{31}{}^{sim}$), $d_{31}$ becomes:

$$d_{31} = \frac{c_{meas}}{c_{sim}} d_{31}^{sim}. \quad (4)$$

This relation holds up as the charge coefficient and the resulting curvature have a linear relationship.

The obtained charge coefficient for different applied electric fields can be seen in figure 3. The maximum deflection of the beams that we observed in the measurements is $\sim$400 $\mu$m for Ekulit-PZT, which corresponds to the strain of $1.8 \times 10^{-4}$. The corresponding variation in the Young's modulus is $\sim$2 GPa (figure 2) which causes a change in our measurement of the transverse charge coefficient of approximately 1 pm V$^{-1}$. Hence, we neglect this change in the measurements as this is smaller than the other error sources of that measurement. The piezo-material AP-PZT856 from American piezo has a thickness of 300 $\mu$m, and we could reliably measure its deflection only at high electric fields. We limited the maximum electric field for each material to the point where an increase in field strength resulted in electrostatic breakdown of the beam or through the air at the side of the beam. Ekulit-PZT ceramics shows the highest $d_{31}$ of $-487$ pm V$^{-1}$ when actuated with an electric field of 1 kV mm$^{-1}$ followed by JM-M1334 and JM-M1100. We will discuss the performance more in detail in section 4, where we also take into account the coercive field strengths.

It turns out that the linear scaling of equation (4) applies in simulations for large scaling factors only if we simultaneously scale the relative permittivity proportional to $d_{31}^2$. I.e. if we want to use the $d_{31}$ measured above to simulate the piezoelectric material, we need to scale the relative permittivity as $\varepsilon_{rnew} = \varepsilon_r \left(\frac{d_{31}}{d_{31}^{PZT5H}}\right)^2$.

To verify that this scaling also applies in nature, we show in figure 4 the relative permittivity depending on the corresponding charge coefficient of JM-M1334, in addition to the best square law fit, $\varepsilon_r = a \, d_{31}^2$ and a quadratic scaling of the standard values in COMSOL. We see that the measured values are, within their uncertainties, in good agreement with a square law, and that this fit also agrees within $3\sigma$ of the scaled built-in values, validating the approach to scale the permittivity quadratically when scaling the $d_{31}$ in a simulation. The measured permittivity showed similar scaling with a mean deviation of 8% from the data found in the literature which is measured within a lower electric field of 0.15–0.45 kV mm$^{-1}$ [23].

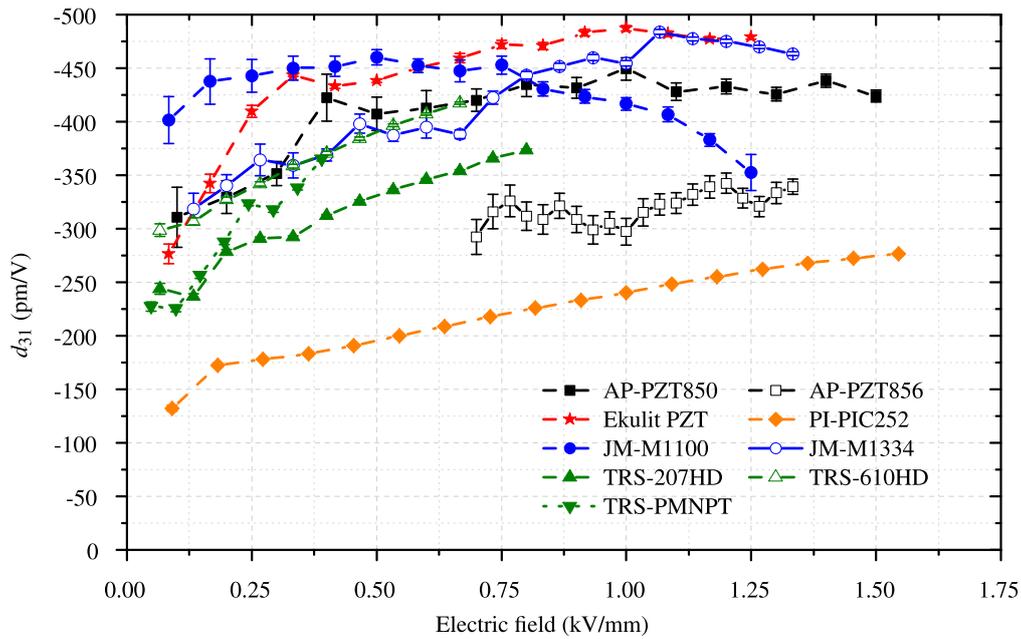

**Figure 3.** The charge coefficient $d_{31}$ of different materials as a function of the electric field strength (E) with their datasheet values for $d_{31}$ marked on the y-axis.

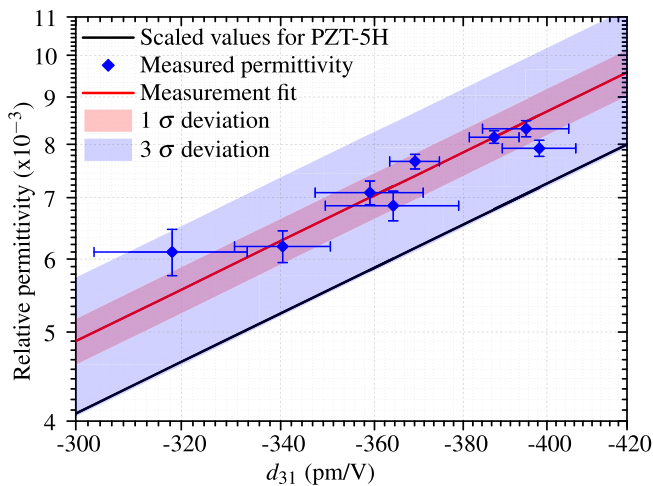

**Figure 4.** The relation between the relative permittivity and the charge coefficient measured at a set of field strengths from 0.1 to 0.6 kV mm$^{-1}$ for JM-M1334 plotted on a double-logarithmic scale. The line show a square law fit and quadratic scaling of standard values.

## 3. Operation region

Below the Curie temperature, the temperature dependence of the strain and the coercive field strength is decisive for the piezoelectric behavior. In this section we measure these properties and in that context also other temperature dependencies. We also investigate the effects of negative operation cycles that approach-$E_c$ and their consequences in long-term operation.

### 3.1. Curie temperature

Above the Curie temperature, ferroelectric ceramics lose their spontaneous polarization. Near the Curie temperature, the dielectric constant $\varepsilon_r$ diverges, and at $T \gtrsim T_c$, it is described by the Curie–Weiss law, in terms of the Curie constant $K_{\text{Curie}}$, temperature $T$ and Curie temperature $T_c$:

$$\epsilon_r = \frac{K_{\text{Curie}}}{T - T_c} + 1. \qquad (5)$$

From an engineering point of view, however, it is more interesting to measure the Curie temperature when approached from below. So we again measured the capacitance across the electrodes of the piezoceramic while ramping the temperature to 300 °C. For continuity in the complex plane, we expect the divergence also to be of 1st order when approached from below, so we fit a generic first order divergence

$$C = a + b\,T + \frac{c}{T_c - T} \qquad (6)$$

as shown in figure 5(a). The fit was done over the region with largest variation in capacitance. The choice of that region causes a deviation of 1.0 °C. We also fitted the usual Curie–Weiss law

$$C = a + \frac{c}{T - T_c} \qquad (7)$$

at $T > T_c$, for materials where we had sufficient data. The variation in the fitting region causes a deviation of 6.3 °C when fitted above $T_c$, and we found good agreement between both methods with an average (rms) deviation of just 8.7 °C.

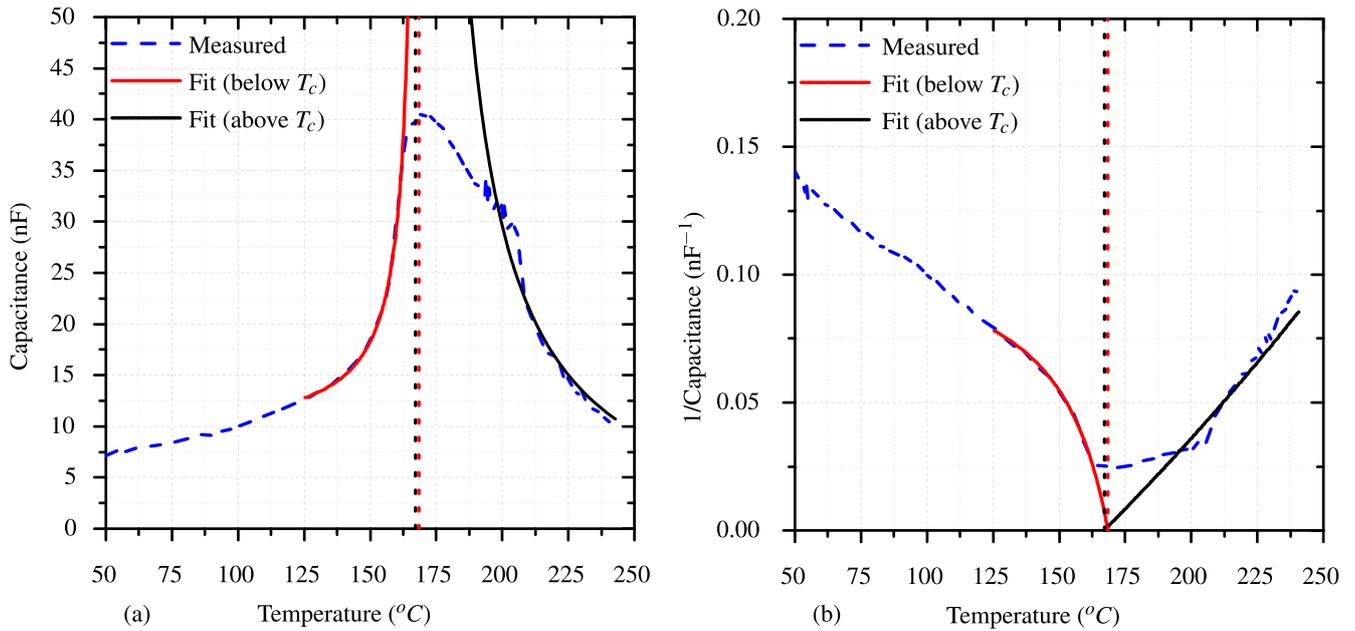

**Figure 5.** The measured change in the capacitance (a) and the inverse of the capacitance (b) of the material (*JMM1334*) with respect to the temperature and the fits for obtaining the Curie temperature above and below $T_c$.

The resulting values for the Curie temperature are given in table 2.

### 3.2. Effect of the temperature on the Young's modulus and charge coefficient

When the temperature increases, the material becomes softer and at the same time, the Weiss domains may become more flexible, affecting also the charge coefficient $d_{31}$. We evaluated this change in Young's modulus depending on the operating temperature by measuring the resonance frequency as mentioned in section 2.3 at different temperatures. As an example, we kept an Ekulit-PZT beam in a mount milled out of aluminum with an attached resistive heater. A temperature controller (*UR484802* from *Wachendorff*) was connected to the resistive heater. A temperature sensor (Pt100) was placed near the beam as the feedback to the controller and the mount was closed using a glass cover. The mount was designed to be filled with oil to achieve a uniform temperature, but this will affect the resonance frequency and also the measurement using the optical profilometer as the refractive index of the oil changes with temperature. Hence, even though the resistive heater was able to achieve up to 90 °C, the temperature values were stable only up to 60 °C.

Similarly, we measured the deflection of the tip of the cantilever in the quasistatic limit at 1 Hz to measure the temperature scaling of the charge coefficient, taking into account the change in the Young's modulus. For both measurements, we assumed the Young's modulus of the glass to be approximately constant as the PZT is assumed to have stronger variations due to the proximity to the Curie temperature. In figure 6 we see an almost linear dependence of the Young's modulus of the material on the temperature with a coefficient $\alpha_Y \sim -0.0014\,\text{K}^{-1}$. The charge coefficient changes linearly with $\alpha_{d_{31}} \sim 0.002\,\text{K}^{-1}$, agreeing well with the values measured by Wang *et al* ($\sim 0.002\,\text{K}^{-1}$) in the same temperature range [24].

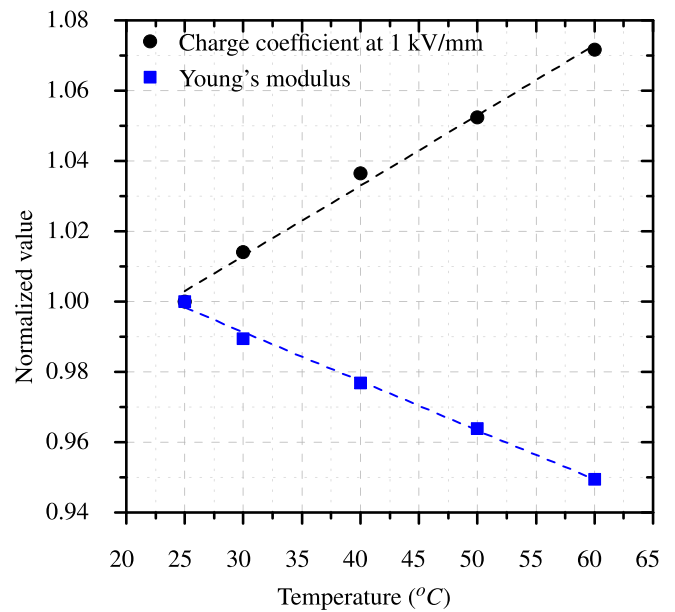

**Figure 6.** The relative temperature scaling of the Young's modulus and the charge coefficient of *Ekulit-PZT*.

### 3.3. Temperature-dependent coercive field strength

The electric current flowing through piezoelectric ceramics when applying an AC signal is a result of aligning Weiss domains inside the ceramic thereby polarizing the material in the direction of the applied electric field; the coercive field is the electric field required to bring the average polarization in the piezoceramic back to zero. To measure the coercive field, we applied a sinusoidal voltage signal with a frequency of

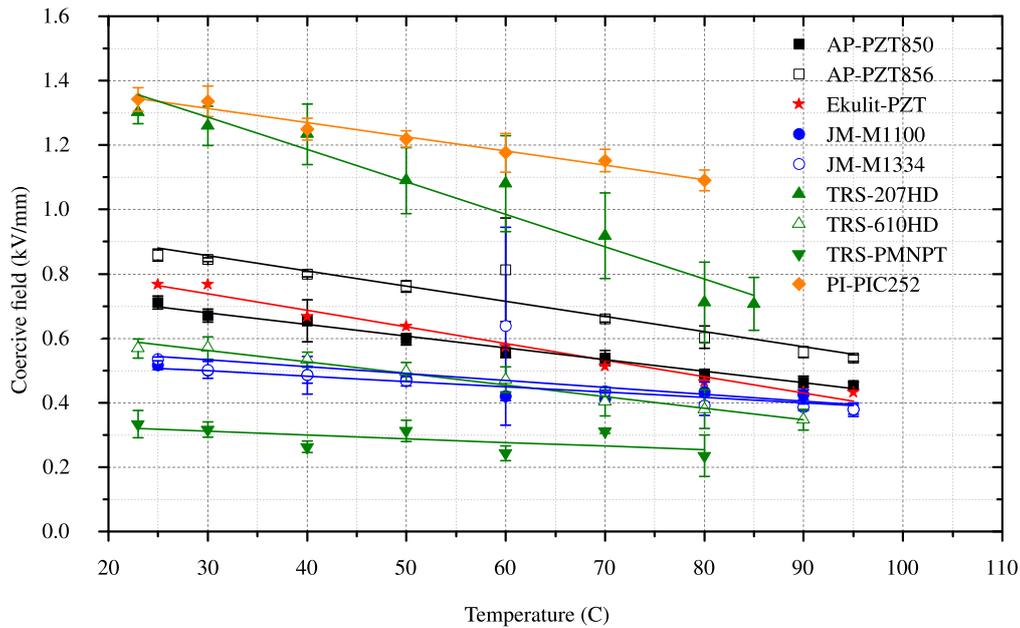

**Figure 7.** The coercive field strength of the different materials as a function of the temperature. The $E_c$ mentioned in the material datasheet is marked on the *y*-axis.

100 mHz and an amplitude that is high enough so that the domain switching can be reliably observed in both in-and-against the direction of original polarization. We kept the sample in the same mount described in section 3.2 this time filled with oil to reduce the risk of electrostatic breakdown due to high electric fields and to optimize the uniformity of the temperature that we varied in steps of 10 °C.

The coercive field is the null point on the polarization curve, which in turn is given by the integrated current. As the pre-polarized piezo films, however, have asymmetric polarization curves, we instead use the voltage at which the piezoceramic material draws the maximum current. The peak current indicates that the domain switching is taking place and this domain switching current is very large compared to the leakage current at the maximum electric field in the ferro-electric materials [28]. Strictly speaking, this gives the field strength of the fastest re-poling but in practice, it will be very close to the coercive field strength. In figure 7 we see that all materials show an approximately linear decrease in the coercive field against the direction of the pre-polarization with increasing temperature.

### 3.4. Electric operation region

The main limitation on the operation of piezoceramics resulting from the coercive field, is the maximum electric field that we can apply against the polarization direction of the piezo materials without risking depolarization. A common assumption has been that the maximum electric field against the direction of polarization should be kept smaller than 33% of the coercive field strength to prevent depolarization [29]. To investigate this limit further, we excited the beam with a 1 Hz signal of different amplitudes and measured the deflection averaged over 5 cycles in two configurations: negative cycles, where the applied electric field is against the direction of polarization and symmetric cycles, where the piezo beams were excited with a symmetric electric field in-and-against the direction of polarization. All the measurements were performed on piezoelectric beams fabricated using *PIC252* from *PI Ceramics*.

Figure 8 shows the mean strain in the piezo obtained from the displacement at the tip of the beam for different voltage cycles. For the negative cycles (figure 8(a)), we see that the piezo unimorph beams were able to operate down to 61% of the coercive field without any significant loss in polarization. When going above 61%, we observe a reduction in the slope of the hysteresis curve and finally the reversal of the polarization direction when the electric field increases above the coercive field. For the symmetric cycles in figure 8(b), we see that the operating region can be extended to approximately 95% of the coercive field against the direction of polarization. When applying field strengths above 95% against the direction of polarization, the material starts to show first depolarizing effects. Even though the symmetric driving at 95% of $E_c$ introduces larger hysteresis in the material, this kind of actuation can be useful for systems with closed loop control or for systems with periodic operation as in the case of a micropump [5, 30].

Given this observation, we introduced a quick re-poling method to improve the working range against the polarization direction, where a voltage pulse is given in the direction of polarization at the end of a negative voltage cycle. The duration of the pulse was kept about 20 times the resonance frequency of the beam with an amplitude of 150% of the coercive field. Figure 9(a) shows the quick re-poling cycle with the pulse at the end. The short duration of the pulse is critical in order to repolarize the domains without giving the material enough time to react mechanically to the pulse. In figure 9(b), we see that it was possible to increase the operating range against the direction of polarization to approximately 75% of the coercive field, above which the slope of the hysteresis curve decreases until the reversal of the polarization direction at 115% of the coercive field. A short resonance can still be observed as shown at one example in figure 9(c) due to the polarizing pulse, but we assume that this will be suppressed in a system with higher damping or by decreasing the pulse width, which was in our case limited by the function generator.

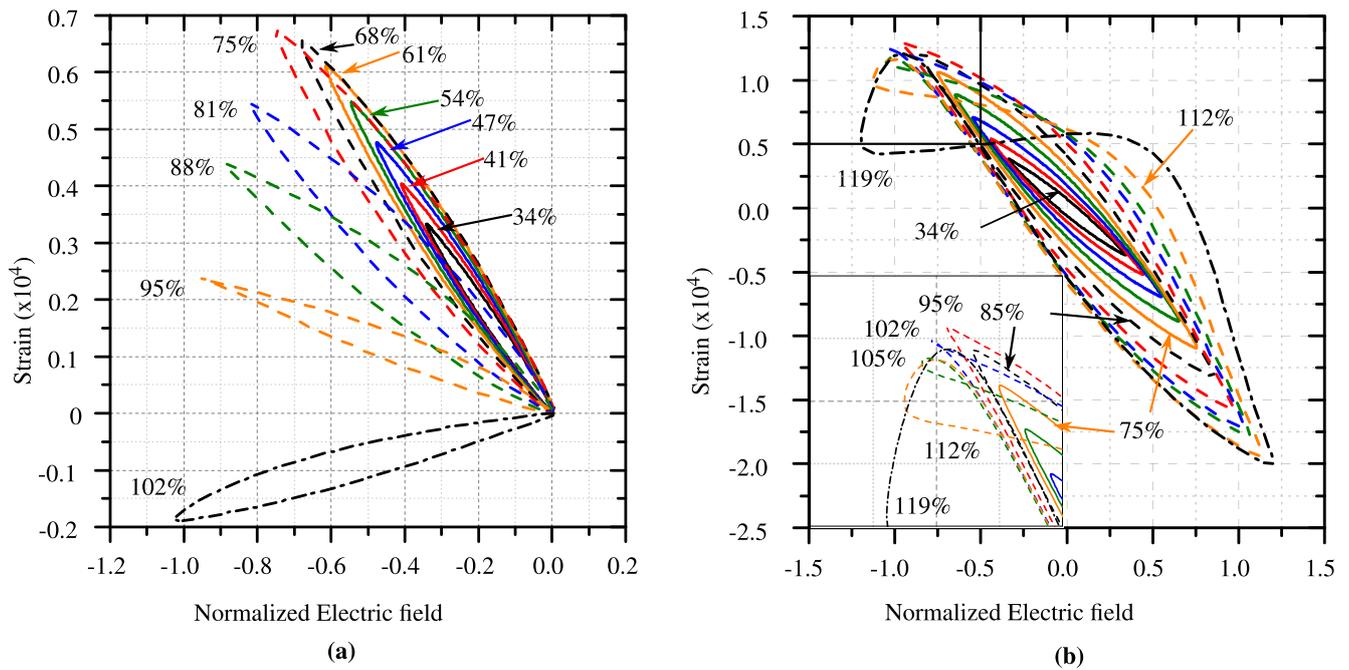

**Figure 8.** Mean strain obtained from the deflection measured at the tip of a piezoelectric bending beam made of *PI-PIC252* at different electric fields when operated in negative only cycles (a) and symmetric cycles (b) up to a certain percentage of the coercive field strength (1.3 kV mm$^{-1}$). The inset in (b) shows the behavior of beam at higher fields against the direction of polarization. The applied electric field on the x-axis is normalized to the coercive field of the material.

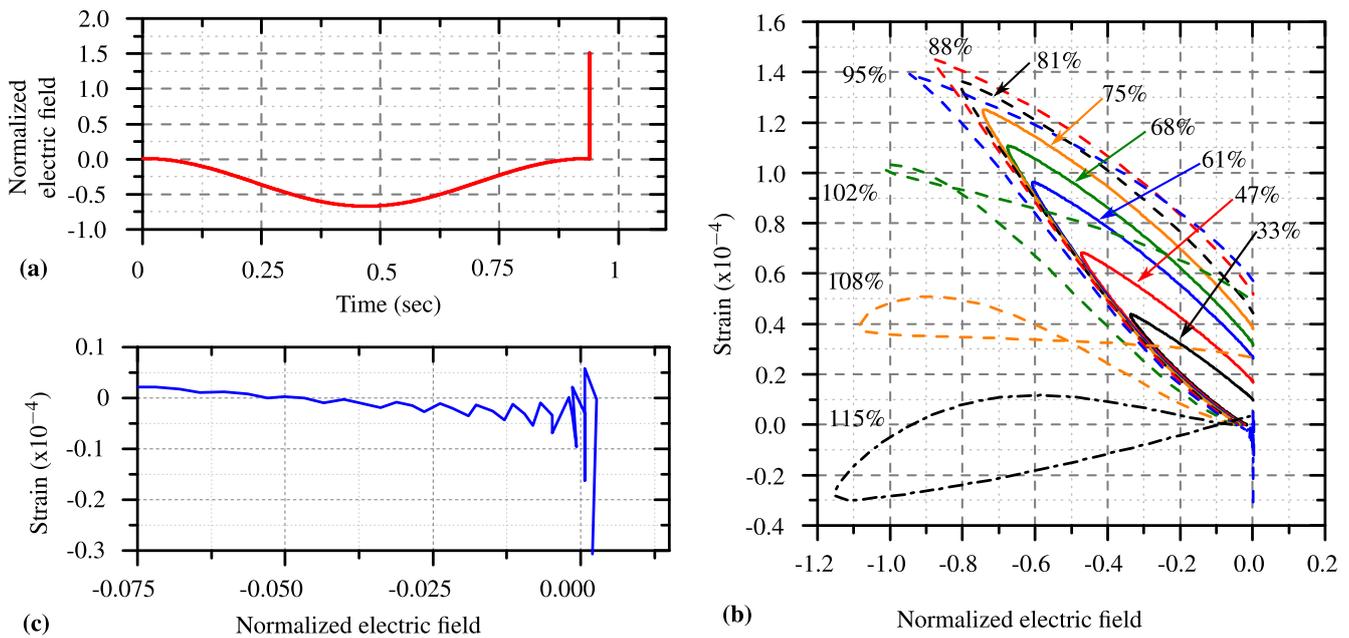

**Figure 9.** (a) Quick re-poling waveform, (b) deflection measured at the tip of the piezoelectric beam made of (*PI-PIC252*) at different electric fields when operated in the quick re-poling mode and (c) zoom-in on the overshoot due to the quick re-poling pulse.

To evaluate the long-term effects of these operation cycles, we actuated the piezoelectric beam in different configurations for 1 million cycles. First, we polarized the piezobeam by applying 150% of the coercive field ($E_c$) at 1 Hz in the direction of polarization for five minutes. Afterwards, we subjected the piezoelectric beams to negative cycles of amplitude 33% of $E_c$ at 10 Hz and measured the deflection at the tip of the beam every five minutes. After operating for one million cycles, we polarized the beam again by applying 150% of $E_c$ and repeated the process for 50%–75% of $E_c$ in negative cycles, then for 66%–95% of $E_c$ applied symmetrically in-and-against the direction of polarization, and finally, 66%–85% of $E_c$ with the quick re-poling method. To evaluate the repeatability of the process, we subjected another beam made from the same material (*PI-PIC252*) to 66% of $E_c$ in symmetric cycles.

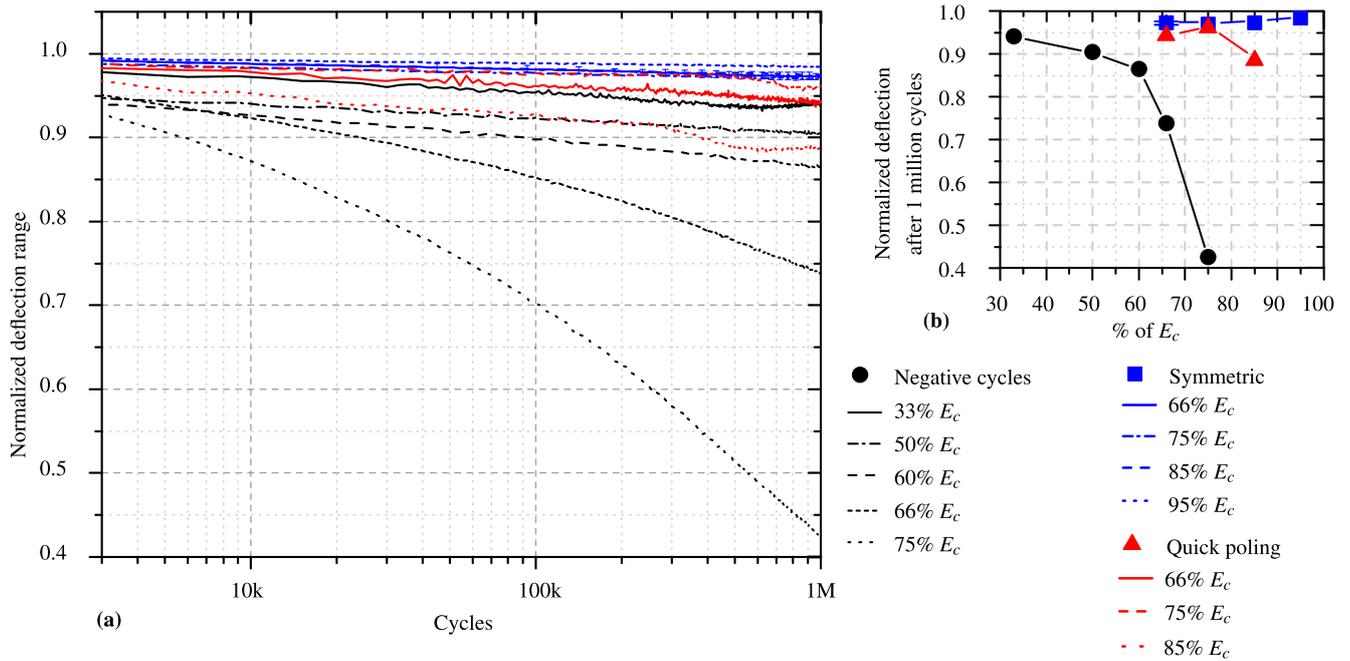

**Figure 10.** (a) Amplitude of the deflection at the tip of a beam made of *PI-PIC252* when operated in different modes, normalized to the initial amplitude after each polarization. (b) Normalized deviation after 1M cycles. The error bars indicate the standard deviation of deflection measured for two different beams.

Figure 10(a) shows the amplitude of the deflection over the 1 million cycles observed every 3000 cycles. We see that, if we apply a maximum of 33% of $E_c$ against the direction of polarization, as suggested by Ruschmeyer [29], the maximum deflection amplitude, and hence the charge coefficient, decreases by 6% after operating 1 million cycles. In comparison, if we apply 66% of $E_c$ in negative cycles, we observe a decrease in the deflection of 26% with no indication of saturation. Symmetric cycles measured up to 95% $E_c$, however, show a decrease in the amplitude of less than 3%. Moreover, applying 95% of $E_c$ was more stable over time than driving with 85% of $E_c$ or less. With the quick re-poling pulse, the beams can be driven with 75% of $E_c$ with a decrease of the amplitude similar to the negative cycles of 33% $E_c$.

## 4. Discussion

The material parameters measured during this study are summarized in tables 2 and 3 along with their datasheet values given in brackets. In section 2, we first presented the measurement of the density and the relative permittivity of the piezoceramics. Both parameters show minimal deviation from the values in the datasheet. The largest difference is seen in the coercive field strength, where, all datasheet values are higher than the measurement results and we observed that the coercive field decreases linearly with increasing temperature. In the measurement of the Young's modulus in section 2.3, we saw that at higher strains, the material becomes softer.

We measured the dependence of the charge coefficient on the applied electric field up to 1.5 kV mm$^{-1}$ in section 2.4. Below 0.15 kV mm$^{-1}$, *JM-M1100* and *PI-PIC252* show a similar rate of change as one of the tip deflection observed by Wang *et al* [17] for soft PZT ceramics. All materials except *TRS-610HD* and *PI-PIC252* started with higher $d_{31}$ values than those mentioned in their datasheets. As seen in figure 3, the measured $d_{31}$ increased with increasing applied electric field and most of them saturated near $E_c$ and then decreased slowly at much higher fields. Table 3 shows the mechanical strain of the materials at different electric fields. The maximum $d_{31}$ measured for each material is compared to the $d_{31}$ value given in the datasheet in figure 12(a). We see that all materials achieve higher $d_{31}$ values than the value mentioned in the datasheet. In our measurement, the maximum $d_{31}$ achieved by *TRS-610HD* was close to the value in the datasheet. *Ekulit-PZT* and *AP-PZT856* have no material datasheet available, so we compared their $d_{31}$ to the standard values for PZT-5H and for *TRS-PMNPT*, we used 40% of the $d_{33}$ value in the datasheet. Taylor *et al* observed that the $d_{31}$ of PMN-PT increases and saturates at about 0.35 kV mm$^{-1}$ and decreases afterwards when measured at 500 Hz [21]. However, in our measurements the $d_{31}$ of PMN-PT increases without saturating within the measured range of electric field. Even though PMN-PT has a high $d_{33}$ on paper, its $d_{31}$ was comparable to the PZT materials and the lower $E_c$ means that it generates less strain compared to the PZT ceramics, making it less desirable for actuator applications.

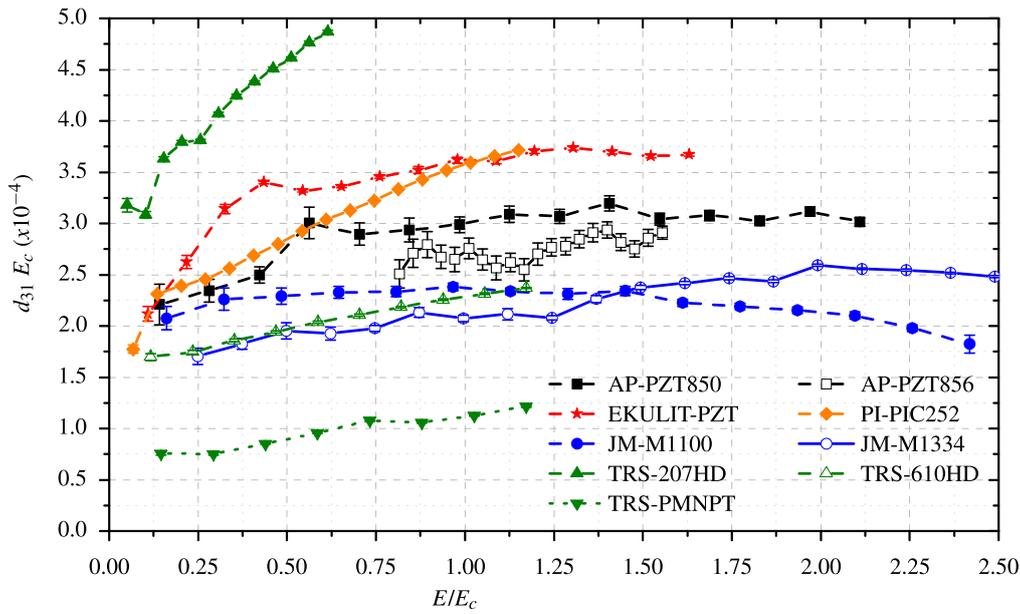

**Figure 11.** The strain normalized to the coercive field strength as a function of the electric field, also normalized to the coercive field of the material. The datasheet values are indicated on the y-axis.

**Table 3.** Young's modulus, strain at different electric field strengths and charge coefficient, the latter compared to the datasheet values.

| Material | Young's modulus (GPa) | Strain | | $d_{31}$ | | |
|---|---|---|---|---|---|---|
| | | At 1 kV mm$^{-1}$ ×10$^{-4}$ | At $E_c$ ×10$^{-4}$ | Maximum (pm V$^{-1}$) | At 0 kV mm$^{-1}$[a] (pm V$^{-1}$) | Datasheet (pm V$^{-1}$) |
| AP-PZT850 | 70 ± 6 | −3.2 ± 0.1 | −3.0 ± 0.1 | −449 ± 9 | −288 ± 13 | (−175) |
| AP-PZT856 | 46 ± 6 | −2.5 ± 0.1 | −2.8 ± 0.1 | −342 ± 9 | −343 ± 75 | (−274)[b] |
| EKULIT-PZT | 37 ± 2 | −3.73 ± 0.02 | −3.60 ± 0.03 | −487 ± 2 | −282 ± 18 | (−274)[b] |
| JM-M1100 | 36 ± 2 | −2.15 ± 0.03 | −2.38 ± 0.04 | −460 ± 7 | −396 ± 8 | (−315) |
| JM-M1334 | 62 ± 6 | −2.44 ± 0.01 | −2.08 ± 0.03 | −483 ± 2 | −290 ± 12 | (−230) |
| TRS-207HD | 55 ± 5 | −5.10 ± 0.01[a] | −5.28 ± 0.02[a] | −373 ± 1 | −219 ± 8 | (−202) |
| TRS-610HD | 68 ± 6 | −2.62 ± 0.01[a] | −2.31 ± 0.01 | −417 ± 1 | −277 ± 3 | (−380) |
| TRS-PMNPT | 27 ± 2 | −1.75 ± 0.01[a] | −1.12 ± 0.01 | −365 ± 2 | −193 ± 17 | (−180) |
| PI-PIC252 | 40 ± 5 | −3.22 ± 0.01 | −3.59 ± 0.01 | −276 ± 1 | −136 ± 5 | (−480)[c] |

[a] Extrapolated data.
[b] Standard value for PZT-5H [31].
[c] 40% of the $d_{33}$ value in the datasheet.

To create a benchmark figure, taking into account that one can adjust the applied field strengths to the material, we introduce normalized charge coefficient $d_{31} E_c$ as a function of the electric field normalized to the coercive field ($E_c$) of each material as shown in figure 11. The piezoceramic TRS-207HD has a high coercive field, so it was not possible to achieve 100% of its coercive field without destroying the beam. Extrapolating the data to $E_c$, however, it suggests the highest achievable strain. In figure 12(b), we compare the force and the strain inside the material at an electric field of 0.75 kV mm$^{-1}$. We further indicate the product $d_{31}^2 Y$ as a measure of the mechanical energy density. Since no $d_{31}$ data is available for PMN-PT at that electric field, the maximum $d_{31}$ measured was taken for comparison.

In section 3, we first measured the Curie temperature. We found that the materials TRS-207HD and PI-PIC252 have the highest Curie temperature and the highest coercive field strength. They also had the lowest charge coefficient but with very smooth and hence predictable behavior. We also observed that there is an approximately linear decrease in the coercive field and the Young's modulus with increasing temperature. If we extrapolate the temperature dependence of $E_c(T)$ to the temperature where $E_c$ vanishes, we find that this temperature lies within 25% of $T_c$ for most of the materials except *TRS-207HD* and *JM-M1100*.

We finally investigated the operating region in terms of the electric field applied. The aim was to find what happens to the material if the electric field against the direction of polarization increases beyond 33% of the coercive field; a limit found in the literature [29]. Figure 13(b) compares the different operating modes at different driving voltages. When driving with negative only cycles, the displacement increases until 75% of $E_c$ and decreases afterward. However, this will result in the reduction in performance of the actuator by 58% over 1 million cycles as seen in figure 10, making it suitable only for short-term operation. We found that operating up to 50% of $E_c$ is more stable for long-term actuation using negative cycles with a reduction by 15% after 1 million cycles. Using symmetric cycles, the maximum displacement increases until the voltage reaches 95% of $E_c$ against the polarization direction and deflection obtained at 95% of $E_c$ was more stable over time than driving with 85% of $E_c$ or less (figure 10). To combine this advantage with purely negative cycles, we introduced a quick re-poling method which increased the operating region against the polarization direction to 85% of $E_c$ as seen in figure 9, by adding a very short positive re-poling pulse at the end of each cycle. In the long-term test of the quick re-poling cycles at an amplitude of 75% of $E_c$, we observed only 6% reduction in performance of the actuator after driving the piezoelectric beam for 1 million cycles (figure 10).

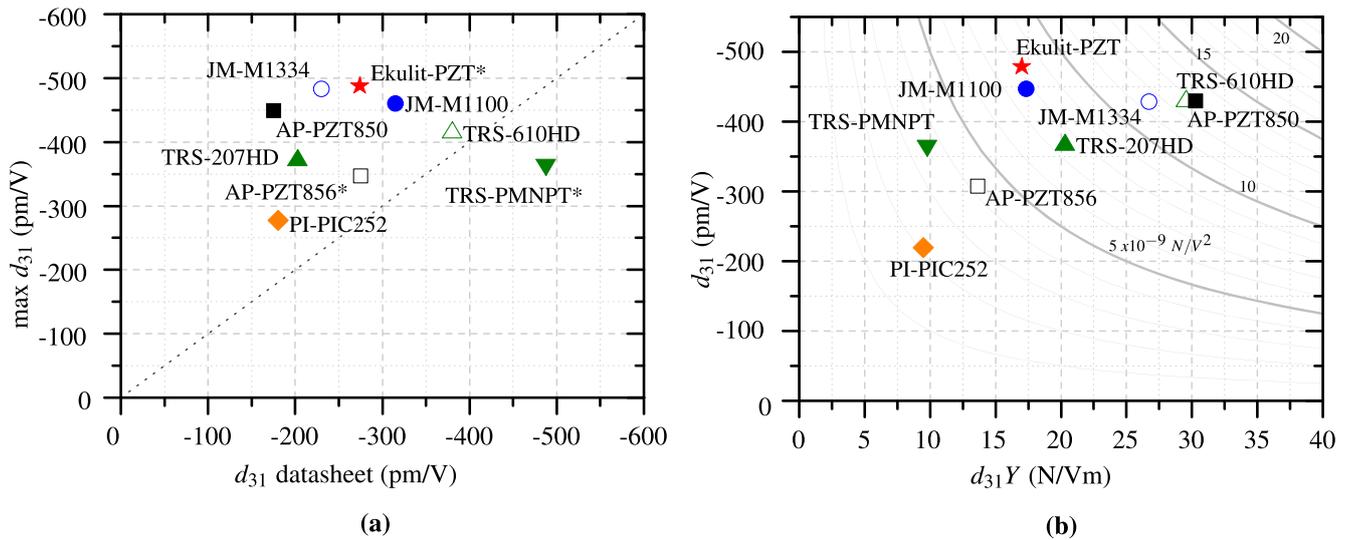

**Figure 12.** (a) Comparison of $d_{31}$ datasheet value and the highest measured value. Since no datasheet is available for Ekulit-PZT and AP-PZT856, the standard $d_{31}$ value of PZT-5H material was assigned as the datasheet value. The $d_{31}$ of TRS-PMNPT was estimated by taking 40% of $d_{33}$ value in the datasheet. (b) The charge coefficient and the product of the charge coefficient and Young's modulus—representing the achievable force—at an electric field of 0.75 kV mm$^{-1}$. The lines indicate constant values of $d_{31}^2 Y$, representing mechanical energy density.

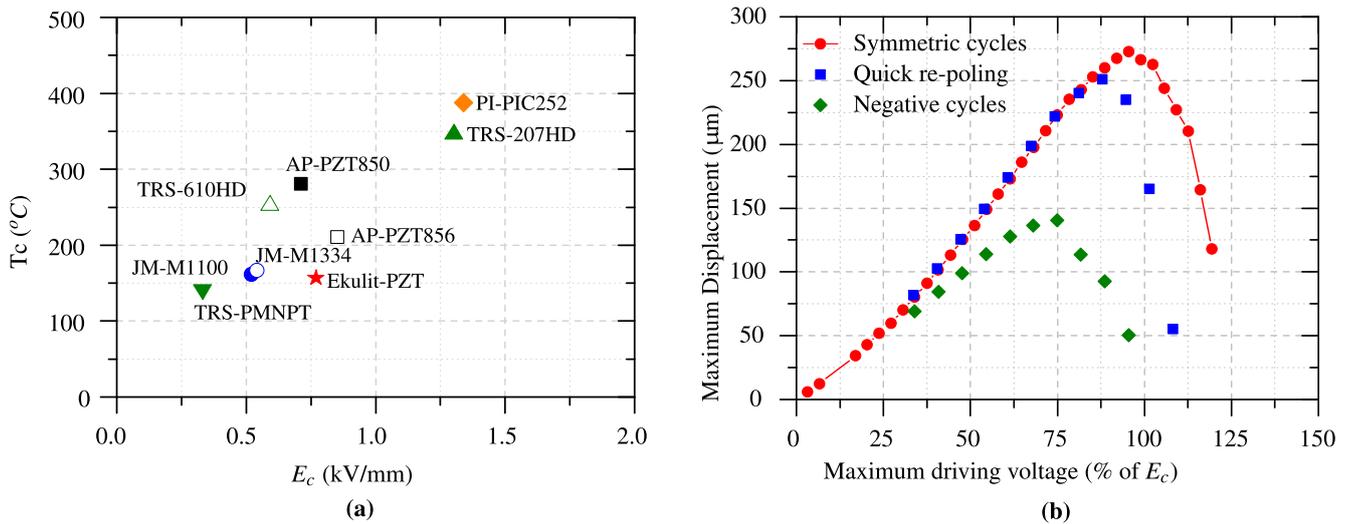

**Figure 13.** (a) The measured Curie temperature $T_c$ and the measured coercive field $E_c$ of the materials. (b) The maximum displacement in the direction against polarization as a function of the driving amplitude represented by % of $E_c$ for different operating modes.

## 5. Conclusions

We studied the nonlinear charge coefficients of piezoceramic materials at high electric field strengths and found in all materials that the charge coefficient increases significantly with increasing fields by up to 109%, until it saturates near the coercive field strength and then slowly starts to drop. Furthermore, we found our measurements overall in good agreements with the literature and are more comprehensive than the literature.

While most FEM simulations cannot handle nonlinearity of piezoceramics, one can use this data to implement a nonlinear simulation by mapping the applied electric field in the simulation to a new corrected value considering the nonlinear $d_{31}$.

Comparing the different materials, we found that the PMN-PT material, despite having a high $d_{33}$ datasheet value, does not show a higher transverse strain in our configuration than PZT materials and is furthermore limited by its low coercive field strength and low Curie temperature. *TRS-207HD* and *PI-PIC252* have by far the highest $T_c$ (350 °C and 389 °C) and $E_c$ (1.3 and 1.34 kV mm$^{-1}$) compared to typical values of 200 °C and 0.6 kV mm$^{-1}$, but a relatively low charge coefficient (373 and 276 pm V$^{-1}$). Taking into account forces, *AP-PZT850*, *TRS-610HD* and *JM-M1334* combine the highest stiffnesses (70, 68, and 62 GPa) with a high charge coefficient. Overall, *AP-PZT850*, *Ekulit-PZT*, *JM-M1100*, *JM-M1334*, and *TRS-610HD* achieve a high charge coefficient between −400 and −500 pm V$^{-1}$—however at different fields. Below 0.75 kV mm$^{-1}$, *JM-M1100* has the highest values and behaves by far most linearly. Above this value, *JM-M1334* and *Ekulit-PZT* have the best performance, while *AP-PZT850* is approximately linear above 0.4 kV mm$^{-1}$.

We furthermore observed that the PZT ceramics can be operated up to 50% of $E_c$ against the direction of polarization and this limit can be extended upto 95% $E_c$ when actuating using symmetric cycles. Furthermore, by using a quick re-poling pulse we were able to expand the electric field limit of negative cyles to 75% $E_c$. In these limits (figure 10) we found less than 10% reduction in performance after one million cycles of operation.

## Acknowledgments

We would like to thank Johnson Matthey Piezo Products GmbH and American Piezo Ceramics Inc. for providing us with samples for our measurements. We extend our gratitude to Professor Thomas Hanemann for theoretical support and we would also like to thank Florian Lemke and Mikel Gorostiaga for their help with the measurement setups and evaluation. This research was supported by Deutsche Forschungsgemeinschaft (DFG) within grant no. WA1657/3-2 and the cluster of excellence BrainLinks-BrainTools, EXC1086.

## ORCID iDs

Binal P Bruno  https://orcid.org/0000-0002-5020-3186